\documentclass[prl,aps,twocolumn,showpacs,superscriptaddress]{revtex4}
\usepackage{amsmath,graphicx}
\pdfoutput=1

\def\prn#1{{\left(#1\right)}}
\begin{document}

\title{Constraints on short-range spin-dependent interactions from scalar spin-spin coupling in deuterated molecular hydrogen} 

\author{M. P. Ledbetter}
\email{micah.ledbetter@gmail.com}
\affiliation{Department of Physics, University of
California at Berkeley, Berkeley, California 94720-7300}
\author{M. V. Romalis}
\affiliation{Department of Physics, Princeton University, Princeton, New Jersey 08544, USA}
\author{D. F. Jackson Kimball}
\affiliation{Department of Physics, California State University --
East Bay, Hayward, California 94542-3084, USA}
%

\date{\today}



\begin{abstract}
A comparison between existing nuclear magnetic resonance measurements and calculations of the scalar spin-spin interaction (J-coupling) in deuterated molecular hydrogen (HD) yields stringent constraints on anomalous spin-dependent potentials between nucleons at the atomic scale (${\rm \sim 1~\AA}$).  The dimensionless coupling constant $g_P^pg_P^{N}/4\pi$ associated with exchange of pseudoscalar (axion-like) bosons between nucleons is constrained to be less than $5\times 10^{-7}$ for boson masses in the range of $5~{\rm keV}$, representing improvement by a factor of 100 over previous constraints. The dimensionless coupling constant $g_A^pg_A^N/4 \pi$ associated with exchange of an axial-vector boson between nucleons is constrained to be $g_A^pg_A^N/4 \pi < 2 \times 10^{-19}$ for bosons of mass $\lesssim 1000~{\rm eV}$, improving constraints at this distance scale by a factor of 100 for proton-proton couplings and more than 8 orders of magnitude for neutron-proton couplings.
\end{abstract}







\maketitle

Over the past few decades, searches for anomalous spin-dependent forces have drawn considerable interest as a signature of the axion \cite{Moo84}, a hypothetical pseudoscalar Goldstone boson \cite{Wei78,Wil78} arising out of the Peccei-Quinn solution to the strong-CP problem \cite{Pec77a,Pec77b}. Axions are also appealing as a candidate for dark matter \cite{Ipser1983}. Other exotic spin-dependent interactions are predicted by a variety of novel theories such as those involving para-photons \cite{Dob05} and unparticles \cite{Geo07}.The possible spin-dependent forces that could arise from exchange of scalar/pseudoscalar or vector gauge bosons are enumerated in Ref. \cite{Dob06}.   Another theoretical framework for considering anomalous spin-dependent interactions comes from introducing non-zero torsion into general relativity, which causes gravity to acquire scalar and vector components that manifest as new spin-mass and spin-spin couplings \cite{Heh76,Sha02,Ham02,Kos08}. Spontaneous Lorentz violation may also generate exotic spin-dependent interactions \cite{Ark05}.

Recent experiments have significantly improved constraints on long-range (tens of cm) dipole-dipole interactions between neutrons \cite{Gle08,Vas09} and monopole-dipole interactions between electrons and nucleons \cite{Hec08}. Earlier work constrained monopole-dipole interactions between nuclei \cite{Ven92} and anomalous dipole-dipole interactions between electrons \cite{Chu93,Bob91} and electrons and nuclei \cite{Win91}. Constraints on new pseudoscalar interactions can also be obtained from spin-independent tests of the inverse square law \cite{Fis99}. Other laboratory experiments constrain anomalous monopole-dipole couplings on length scales in the range of $\mu$m to mm \cite{Baessler2007,Hoe11,Petukhov2010,Fu2011}. Spin relaxation studies of hyperpolarized ${\rm ^3He}$ gas have been used to limit anomalous dipole-dipole interactions between neutrons at length scales of about 100 nm \cite{Fu2012}. Because of the possibility that new force-mediating bosons may be massive and have limited range, it is of considerable interest to find experimental techniques to search for anomalous spin-dependent interactions at even shorter distances.

Here we discuss constraints on the existence of anomalous dipole-dipole forces on angstrom length scales.  Our constraints are obtained by comparison of nuclear magnetic resonance (NMR) measurements and theoretical calculations of J-coupling in deuterated molecular hydrogen (HD).  Such couplings have the form  $J\mathbf{I}\cdot\mathbf{S}$ (here $\mathbf{I}$ and $\mathbf{S}$ are nuclear spin operators) and arise due to a second-order hyperfine interaction. To our knowledge, the only constraints on anomalous nuclear dipole-dipole couplings at these length scales come from Ramsey's molecular-beam measurements of ${\rm{H_2}}$ \cite{Ram79} and studies of spin-exchange collisions between $^3$He and Na \cite{Kim10}.  Atomic-scale constraints on anomalous, dipole-dipole couplings between electrons and nuclei, mediated by exchange of axial-vector bosons, have recently been obtained from the hyperfine structure of hydrogen-like atoms \cite{Kar10,Kar11a,Kar11b}.  The constraints derived in this letter represent a factor of $10^2$ to $10^8$ improvement over previously derived laboratory limits on anomalous spin-dependent forces at the ${\rm \AA}$ range, corresponding to a particle with keV-scale mass.

There are several additional noteworthy features of the analysis presented here.  Constraining monopole-dipole interactions is appealing because such couplings violate invariance under both time reversal $(T)$ and spatial inversion $(P)$, and hence one expects negligible background from standard model physics.  Dipole-dipole couplings, on the other hand, are even under both $T$ and $P$, and arise from standard model physics.  In this sense, dipole-dipole couplings may appear less attractive in searches for exotic physics because one must carefully calculate the effects of standard model physics. Despite this difficulty, our analysis results in constraints that are within two orders of magnitude of hadronic axion models.  This is significantly closer than limits on monopole-dipole couplings come to constraining QCD axions \cite{Baessler2007,Hoe11,Petukhov2010,Fu2011}.  Furthermore, axion mediated  dipole-dipole coupling scales as $1/f_a^4$, ($f_a$ is the Peccei-Quinn symmetry breaking parameter) independent of the QCD $\theta$ parameter.  In contrast monopole-dipole and monopole-monopole coupling depend on $\theta$ as $\theta/f_a^3$ and $\theta^2/f_a^2$, respectively \cite{Moo84}. Therefore, interpreting limits on monopole-dipole or monopole-monopole couplings as a constraint on $f_a$ comes with large uncertainty related to the size of the $\theta$ parameter.   In many discussions, $\theta$ is taken to be order of $10^{-10}$ based on limits from the neutron electric dipole moment experiments, however it may be much smaller, relaxing the constraints one can place on $f_a$.

We also note that stringent constraints on axion couplings can be obtained from astrophysical observations.  Electron-axion or photon-axion couplings can be constrained by cooling rates of low-mass stars \cite{Dic78,Raf95}, requiring the axion mass be less than $10^{-2}~{\rm eV}$.  More relevant in the present context are the constraints obtained from supernova SN1987A \cite{Engel1990} or the metallicity of red giant stars \cite{Haxton1991}, which limit axion-nucleon interactions.  The 14.4 keV emission line of ${\rm ^{57}Fe}$ also constrains axion-nucleon couplings \cite{Derbin2009}.  However, these astrophysical constraints are somewhat particular to the axion, and do not apply to other pseudoscalar bosons or axial-vector interactions \cite{Dob06,Raf07}.

Moody and Wilczek discuss the potentials arising from the exchange of pseudoscalar ($P$) axion-like particles \cite{Moo84} in the non-relativistic limit.  The dipole-dipole potential has the form
\begin{widetext}
\begin{equation}\label{Eq:V3}
V_3(r)=\frac{g_P^1g_P^2}{16\pi M_1M_2}e^{-mr} \left\lbrack\left\lbrace\mathbf{\hat{\sigma}}_1\cdot\mathbf{\hat{\sigma}}_2-3(\mathbf{\hat{\sigma}}_1\cdot\mathbf{\hat{r}})(\mathbf{\hat{\sigma}}_2\cdot\mathbf{\hat{r}})\right\rbrace
\left(\frac{m}{r^2}+\frac{1}{r^3}\right)-(\mathbf{\hat{\sigma}}_1\cdot\mathbf{\hat{r}})(\mathbf{\hat{\sigma}}_2\cdot\mathbf{\hat{r}})\frac{m^2}{r}\right\rbrack.
\end{equation}
\end{widetext}
%
Here, we work in units where $\hbar = c = 1$, $m$ is the mass of the axion-like particle, $g_P^1 g_P^2/\prn{ 4 \pi  }$ is the dimensionless pseudoscalar coupling constant between the particles, $M_1$ and $M_2$ are their respective masses, and $r$ is the distance (in units of inverse energy) separating the two particles.  The subscript on the left-hand side is chosen to match the notation of Ref.~\cite{Dob06}.  We also note that a $\delta$ function contribution to $V_3(r)$ is neglected here because of Coulomb repulsion of the two nuclei.

The measurements from which we extract our constraints occur in gas phase, in which the internuclear vector $\mathbf{\hat{r}}$ suffers random reorientation due to collisions, leading to averaging of Eq. \eqref{Eq:V3}.  The first term in braces has the same angular dependence as the usual magnetic dipolar interaction and averages to zero. After averaging, the second term is proportional to $\mathbf{\hat{\sigma}}_1\cdot\mathbf{\hat{\sigma}}_2$, yielding an effective anomalous J-coupling $\Delta J_3\mathbf{I}\cdot\mathbf{S}$, where $\mathbf{I}$ and $\mathbf{S}$ are the respective spins of the proton and deuteron, with
\begin{equation}\label{Eq:V3ave}
 \Delta J_3 = \frac{g_P^pg_P^D}{4\pi}\frac{1}{2 M_p^2}\frac{m^2e^{-mr}}{3r}.
\end{equation}
Here, $g_P^D = g_P^n+g_P^p$, we have made the approximation that the neutron and proton masses are equal, $M_n = M_p$, and we assume that the proton and neutron of the deuteron each contribute roughly equally to the spin, $\mathbf{S}$, of the deuteron, $\langle\mathbf{\hat{\sigma}}_p\rangle =\langle\mathbf{\hat{\sigma}}_n\rangle = \mathbf{S}$.

Two measurements of $J$ in HD can be found in the literature:
Ref. \cite{Beckett}  reports $J= 42.94\pm 0.04~{\rm Hz}$ and Ref.
\cite{Neronov1975} reports $J= 43.11\pm 0.02~{\rm Hz}$ at 40 K. The mean and standard deviation of these measurements is 43.025 and 0.12 Hz, respectively.
Using density functional theory, Vahtras \textit{et al.} \cite{Vahtras1992} calculate a spin-spin coupling constant of
$J=43.15~{\rm Hz}$ at 40 K, in agreement with measurements at the level of 0.12 Hz.  No estimate for the uncertainty of the calculation was reported.  We take this to constrain $\Delta J_3 < 0.24~{\rm Hz}$ (${\rm 9.8\times 10^{-16}~{\rm eV}}$) at the 2-sigma level. The distance between
proton and deuteron is about $r=1.4$ Bohr radii \cite{Vahtras1992}, or in units where $\hbar = c = 1$, $r = 0.00038~{\rm eV^{-1}}$.

\begin{figure}
\center
\includegraphics[width=3.35 in]{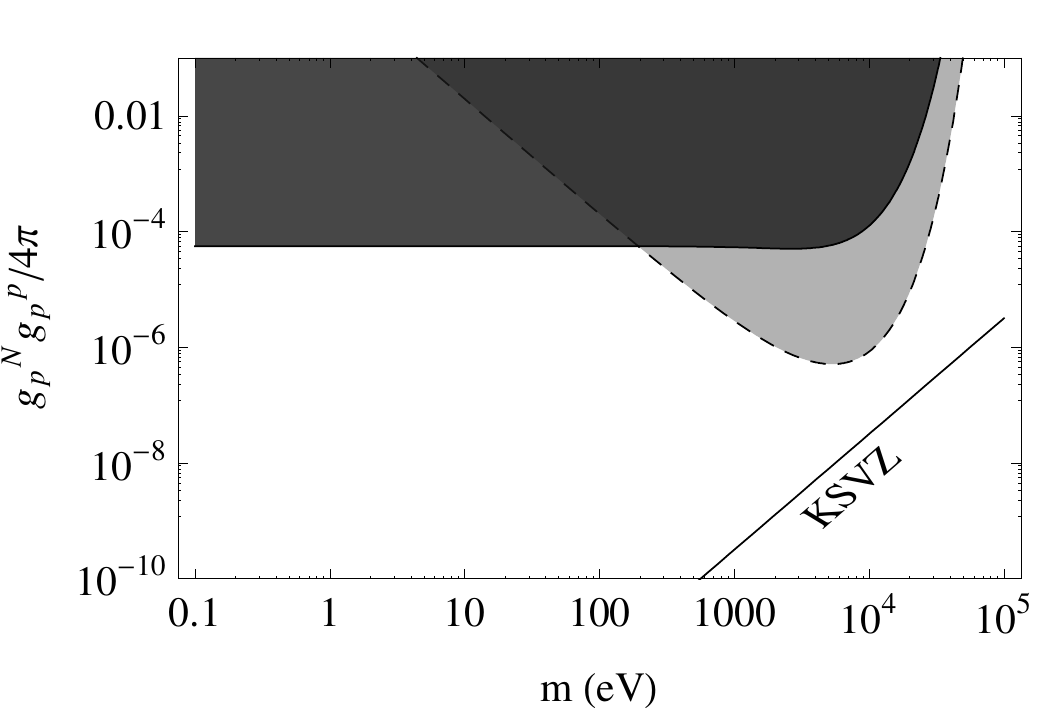}
\caption{Constraint (at the 2-$\sigma$ level) on the dimensionless coupling constant $g_P^N g_P^p/4 \pi$ associated with the dipole-dipole potential, $V_3(r)$, as a function of the exchange boson mass, obtained from comparison of measured and calculated J-coupling parameters for HD (dashed line, light gray fill), where $g_P^N=g_P^n+G_P^p$. Also shown is the existing constraint on $p-p$ pseudoscalar couplings from Ref.~\cite{Ram79} (solid line, darker fill). The straight line represents axion coupling in the KSVZ model, given by Eq. \eqref{Eq:KSVZconstraint}.}\label{Fig:HDpseudo}
\end{figure}

From the experimental measurements, theoretical calculations, and Eq. \eqref{Eq:V3ave}, we can constrain the dimensionless coupling parameter $g_P^Dg_P^p/4\pi$ as a function of the pseudoscalar mass $m$, as indicated by the light shaded region bounded by the dashed line in Fig. \ref{Fig:HDpseudo}.  The strongest constraint occurs for bosons with mass $m = 2/r=5300~{\rm eV}$ for which $g_P^Dg_P^p/4\pi<5\times 10^{-7}$.
For comparison, the darker shaded region bounded by the dashed line shows the limits obtained from Ramsey's molecular beam measurements \cite{Ram79} of HH dipole-dipole interactions, where non-magnetic contributions are constrained to be less than $K_{NM}<70~{\rm Hz}$, $( 3\times 10^{-13}~{\rm eV})$, at the $2-\sigma$ level.  The form of the interaction Ramsey considered differs from Eq. \eqref{Eq:V3} in its angular dependence, however it can be approximately interpreted as
\begin{equation}
K_{NM} = \frac{(g_P^p)^2}{4\pi}\frac{1}{4 M_p^2}\left(\frac{1}{r^3}+\frac{m}{r^2}+\frac{m^2}{r}\right)e^{-mr}.
\end{equation}
This limit ($2-\sigma$ level) is indicated by the dark shaded region bounded by the solid line in Fig. \ref{Fig:HDpseudo}.


We note that our constraints are within two orders of magnitude of standard axion models.    The axion mass is determined by the Peccei-Quinn symmetry breaking scale $f_a$
\begin{equation}\label{Eq:ax1}
m_a = \frac{\sqrt{z}}{1+z}\frac{f_\pi m_\pi}{f_a},
\end{equation}
where $f_\pi = 92~{\rm MeV}$ is the pion decay constant, $m_\pi = 134~{\rm MeV}$ is the pion mass, and $z = m_u/m_d=0.56$ is the ratio of up and down quark masses. In the Kim-Shifman-Vainshtain-Zakharov (KSVZ) model (see, e.g. Ref. \cite{Raffelt2008} for a brief review), axions couple only to hadrons but not quarks or leptons at tree level. Its coupling to a nucleon of mass $M_N$ is given by $g_P^N = C_N M_N/f_a$, where $C_N$ is a model-dependent numerical factor. Using expressions for $C_N$ found in Ref. \cite{Raffelt2008} for KSVZ axions, we find
\begin{equation}\label{Eq:KSVZconstraint}
\frac{(g_a^p+g_a^n)g_a^p}{4\pi} = 3.06\times 10^{-16} ({\rm eV^2})m_a^2.
\end{equation}
This relation is given by the straight line in Fig. \ref{Fig:HDpseudo}.  Our analysis, obtained from purely low energy laboratory measurements, is thus within two orders of magnitude of constraining KSVZ axions.

It is worth discussing axion constraints obtained from high energy experiments in the context of our limits.  To our knowledge, the best limits come from the decay of the $K^+$ particle. Reference \cite{Adler2002} limits the branching ratio ${\rm BR}(K^+\rightarrow \pi^++a) < 4.5\times 10^{-11}$.  The relation between the branching ratio and the Peccei-Quinn symmetry breaking scale can be found in Ref. \cite{Georgi1986}:
\begin{equation}
f_a^{-1} = \sqrt{\frac{\rm BR}{5.6 f_\pi^2}\left(1+\frac{2}{z}+\frac{1}{z^2}\right)}.
\end{equation}
Using values of $C_N$ found in Ref. \cite{Raffelt2008}, and for $z = 0.56$, we find $f_a > 11600~{\rm GeV}$ (corresponding to $m_a<507 ~{\rm eV}$), and interpreted in the context of Fig. \ref{Fig:HDpseudo}
\begin{equation}
\frac{(g_a^p+g_a^n)g_a^n}{4\pi} < 9\times 10^{-11}.
\end{equation}
While these high energy limits are significantly stronger than those obtained from HD, the latter are specifically sensitive to the spin-dependence of the interaction.

\begin{figure}
\center
\includegraphics[width=3.35 in]{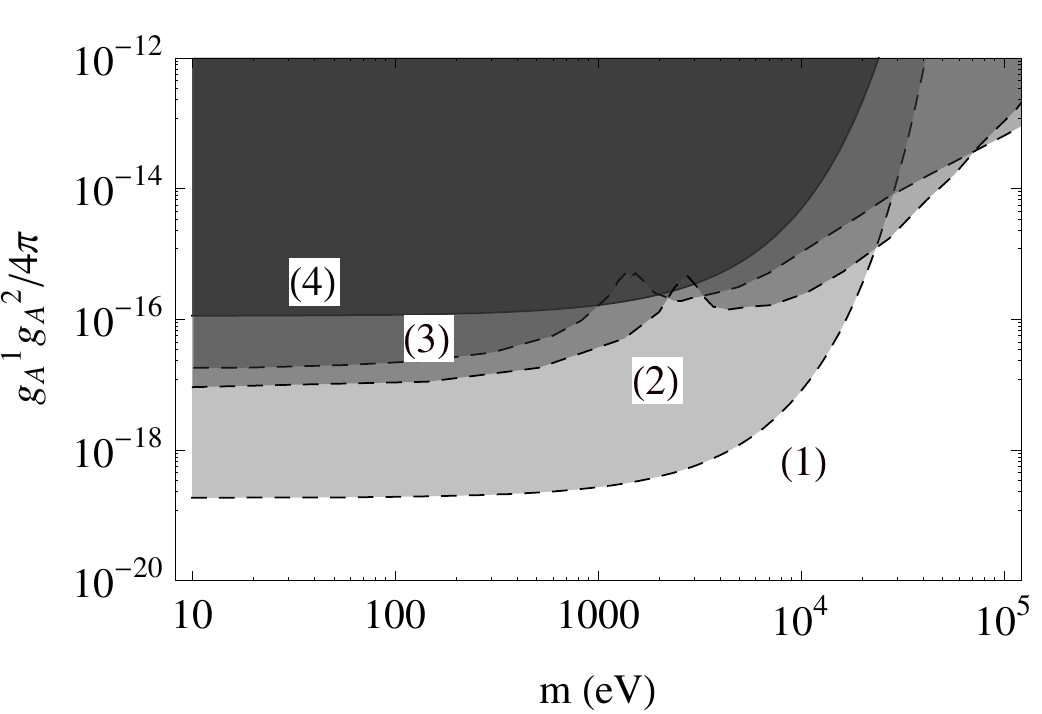}
\caption{Constraints (at the 2-$\sigma$ level) on the dimensionless coupling constants $g_A^1 g_A^2/\prn{ 4 \pi }$ as a function of the mass of an axial vector bosons ($A$) between nucleons [$V_2(r)$, see Eq.~\eqref{Eq:V2}].  Curve (1) shows the constraints obtained from the J-coupling interaction in HD, where $1=p$ and $g_A^2=g_A^n+g_A^p$.  Curves (2) and (3) show the respective constraints on $e-n$ and $e-p$ axial vector couplings from Ref.~\cite{Kar11a}. Curve (4) shows the constraints obtained from Ramsey's measurements of proton-proton dipole-dipole coupling in molecular ${\rm H_2}$. }\label{Fig:HDlimits-axial-vector}
\end{figure}

In addition to pseudoscalar axion-like particles, the presently considered set of measurements and calculations can be used to constrain spin-dependent forces due to exchange of other particles.  In the case of spin-1 axial vector ($A$) bosons, there arises a Yukawa potential \cite{Dob06}
\begin{align}
V_2(r)  = \frac{g_A^1 g_A^2}{4 \pi }  \frac{1}{r} \mathbf{\hat{\sigma}}_1 \cdot \mathbf{\hat{\sigma}}_2 e^{ - m r  }. \label{Eq:V2}
\end{align}
Again assuming that the nucleons in deuterium contribute equally to its spin, we find that in the limit of massless bosons $(g_A^p+g_A^n)g^p_A/4\pi<Jr/2=1.8\times 10^{-19}$ at the $2-\sigma$ level.  Constraints as a function of boson mass are shown in Fig. \ref{Fig:HDlimits-axial-vector} (curve 1), along with constraints obtained from Ramsey's measurements in ${\rm H_2}$ (curve 4).  We also show the spin-dependent constraints between electron and neutron (curve 2) or proton (curve 3), obtained from measurements and calculations of the $2s$ hyperfine state in atomic hydrogen and deuterium \cite{Kar11a}.  The limits obtained from spin-exchange cross sections in ${\rm ^3He}$-Ne collisions \cite{Kim10} are about eight orders of magnitude worse than obtained in the case of HD J-coupling.

Finally, we point out that J-coupling measurements in an isotropic liquid or gas, where molecules rapidly rotate, are most sensitive to exchange of axion-like particles with masses in the range of $2/r$ [see Eq. \eqref{Eq:V3ave}].  If the molecule is placed in an anisotropic environment such as a liquid crystal, tumbling is inhibited and the molecule becomes partially oriented, thereby rendering the term in curly brackets in Eq. \eqref{Eq:V3} observable.  This term is largest for pseudoscalar exchange boson mass $m=0$. Therefore, precise calculations and measurements of dipole-dipole couplings in liquid crystal environments may yield improved constraints on light pseudoscalar bosons.  Accurate measurements and calculations of dipole-dipole couplings in partially oriented benzene exist \cite{Vaara2002}, however the limits one may extract from them have potentially larger systematic uncertainty and will be discussed elsewhere.


In conclusion, we have used measurements and theoretical calculations of J-coupling in deuterated molecular hydrogen to constrain spin-dependent forces due to exchange of exotic pseudoscalar and axial-vector particles.  This analysis improves by over two orders of magnitude the constraints obtained by Ramsey's comparison of experiment and theory on the dipole-dipole coupling of protons in molecular hydrogen.  In particular the constraints on the exchange of axion-like pseudoscalar particles, is improved by a factor of $10^2$ in the mass range of $10^2-10^4$ eV.  We also discussed the limits these measurements can place on the exchange of photon-like lightweight axial-vector bosons, improving constraints by eight orders of magnitude over limits placed on the proton-neutron couplings, compared to an analysis of spin-exchange in sodium and neon.  New experimental techniques to measure scalar couplings with extremely high accuracy based on ultra-low field NMR \cite{Ledbetter2009} could lead to improvements in experimental precision.  Combined with improved density functional theory calculations, comparison of scalar or dipole-dipole couplings may be used to further constrain spin-dependent forces.  An improvement in the pseudoscalar coupling limit by two orders of magnitude would bring these constraints into the range of standard axion models, providing a new set of purely laboratory limits on the QCD axion in the keV range.
In order to substantially improve these limits it is desirable to find spin systems with a naturally small $J$ coupling to reduce the requirements for precision calculations of standard model interactions. In this regard it is interesting to consider chemically unbound spin systems.  In this case, a finite $J$ coupling can still arise due to second-order hyperfine interaction in van der Waals molecules \cite{Bagno2003}. For example, in a mixture of liquid ${\rm ^{129}Xe}$ and pentane, a J coupling of 2.7 Hz was measured in agreement with predictions \cite{Ledbetter2012}. Lighter atoms would have even smaller J coupling \cite{Pecul2008}. A liquid ${\rm H_2-^3He}$ mixture \cite{Hiza1981} may be interesting to consider in this context. Hyperpolarizing ${\rm ^3He}$ in the mixture would allow one to measure very small frequency shift of H NMR using the techniques described in \cite{Heckman2003, Ledbetter2012}.

\acknowledgments

This work has been supported by grants PHY-0969666 (D.F. Jackson-Kimball), CHE-0957655 (M. P. Ledbetter), and PHY-0969862 (M.V. Romalis) from the National Science Foundation.  The authors thank Dmitry Budker for valuable discussions.


\begin{thebibliography}{99}

\bibitem{Moo84} J. E. Moody and F. Wilczek, Phys. Rev. D {\textbf{30}}, 130 (1984).

\bibitem{Wei78} S. Weinberg, Phys. Rev. Lett. {\textbf{40}}, 223 (1978).

\bibitem{Wil78} F. Wilczek, Phys. Rev. Lett. {\textbf{40}}, 279 (1978).

\bibitem{Pec77a} R. D. Peccei and H. R. Quinn, Phys. Rev. Lett. {\textbf{38}}, 1440 (1977).

\bibitem{Pec77b} R. D. Peccei and H. R. Quinn, Phys. Rev. D {\textbf{16}}, 1791 (1977).

\bibitem{Ipser1983} J. Ipser and P. Sikivie, Phys. Rev. Lett. \textbf{50}, 925-927 (1983).

\bibitem{Dob05} B. A. Dobrescu, Phys. Rev. Lett. {\textbf{94}}, 151802 (2005).

\bibitem{Geo07} H. Georgi, Phys. Rev. Lett. {\textbf{98}}, 221601 (2007).

\bibitem{Dob06} B. A. Dobrescu and I. Mocioiu, J. High Energy Phys. {\textbf{11}}, 5 (2006).

\bibitem{Heh76} F. W. Hehl, P. von der Heyde, G. D. Kerlick, and J. M. Nester, Rev. Mod. Phys. {\bf{48}}, 393 (1976).

\bibitem{Sha02} I. L. Shapiro, Phys. Rep. {\textbf{357}}, 113 (2002).

\bibitem{Ham02} R. T. Hammond, Rep. Prog. Phys. {\textbf{65}}, 599 (2002).

\bibitem{Kos08} V. A. Kostelecky, N. Russell, and J. D. Tasson, Phys. Rev. Lett. {\textbf{100}}, 111102 (2008).

\bibitem{Ark05} N. Arkani-Hamed, H. Cheng, M. Luty and J. Thaler, J. High Energy Phys. {\textbf{7}}, 29 (2005).

\bibitem{Gle08} A. G. Glenday, C. E. Cramer, D. F. Phillips, and R. L. Walsworth, Phys. Rev. Lett. {\textbf{101}}, 261801 (2008).

\bibitem{Vas09} G. Vasilakis, J. M. Brown, T. W. Kornack, and M. V. Romalis, Phys. Rev. Lett. {\textbf{103}}, 261801 (2009).

\bibitem{Hec08} B. R. Heckel, E. G. Adelberger, C. E. Cramer, T. S. Cook, S. Schlamminger, and U. Schmidt, Phys. Rev. D {\textbf{78}}, 092006 (2008).

\bibitem{Ven92} B. J. Venema, P. K. Majumder, S. K. Lamoreaux, B. R. Heckel, and E. N. Fortson, Phys. Rev. Lett. {\textbf{68}}, 135 (1992).

\bibitem{Chu93} T. C. P. Chui and W. T. Ni, Phys. Rev. Lett. 71, 3247 (1993).

\bibitem{Bob91} V. F. Bobrakov et al., JETP Lett. 53, 294 (1991).

\bibitem{Win91} D. J. Wineland, J. J. Bollinger, D. J. Heinzen, W. M. Itano, and M. G. Raizen, Phys. Rev. Lett. {\textbf{67}}, 1735 (1991).

\bibitem{Fis99} E. Fischbach and D. E. Krause, Phys. Rev. Lett. {\textbf{82}}, 4753 (1999); E. Fischbach and D. E. Krause, Phys. Rev. Lett. {\textbf{83}}, 3593 (1999).

\bibitem{Baessler2007} S. Baessler, V. V. Nesvizhevsky, K. V. Protasov, and A. Yu. Voronin, Phys. Rev. D {\textbf{75}}, 075006 (2007).

\bibitem{Hoe11} S. A. Hoedl, F. Fleischer, E.G. Adelberger, and B. R. Heckel, Phys. Rev. Lett. {\textbf{106}}, 041801 (2011).

\bibitem{Petukhov2010} A. K. Petukhov, G. Pignol, D. Jullien, and K. H. Andersen, Phys. Rev. Lett. 105, 170401 (2010).

\bibitem{Fu2011} C. B. Fu, T. R. Gentile, and W. M. Snow, Phys. Rev. D 83, 031504(R) (2011).

\bibitem{Fu2012} C. B. Fu and W. M. Snow, arxiv:1103.0659 (2012).

\bibitem{Ram79} N. F. Ramsey, Physica {\textbf{96A}}, 285 (1979).


\bibitem{Kim10} D. F. Jackson Kimball, A. Boyd, and D. Budker, Phys. Rev. A {\textbf{82}}, 062714 (2010).

\bibitem{Kar10} S. G. Karshenboim, Phys. Rev. Lett. {\textbf{104}}, 220406 (2010); Phys. Rev. D {\textbf{82}}, 073003 (2010); Phys. Rev. D {\textbf{82}}, 113013 (2010).

\bibitem{Kar11a} S. G. Karshenboim, Phys. Rev. A {\textbf{83}}, 062119 (2011).

\bibitem{Kar11b} S. G. Karshenboim and V. V. Flambaum, Phys. Rev. A {\textbf{84}}, 064502 (2011).

\bibitem{Dic78} D. A. Dicus, E. W. Kolb, V. L. Teplitz, and R. V. Wagoner, Phys. Rev. D {\textbf{18}}, 1829 (1978).

\bibitem{Raf95} G. G. Raffelt and A. Weiss, Phys. Rev. D {\textbf{51}}, 1495 (1995).

\bibitem{Engel1990} J. Engel, D. Seckel, and A. C. Hayes, Phys. Rev. Lett. \textbf{65}, 960 (1990).

\bibitem{Haxton1991} W. C. Haxton and K. Y. Lee, Phys. Rev. Lett. \textbf{66} (1991).

\bibitem{Derbin2009} A. V. Derbin \textit{et. al.}, Eur. Phys. J. C \textbf{62}, 755 (2009).

\bibitem{Raf07} G. G. Raffelt, J. Phys. A {\textbf{40}}, 6607 (2007).

\bibitem{Beckett} J. R. Beckett, Ph.D. thesis, Rutgers University (1979).

\bibitem{Neronov1975} Yu. I. Neronov and A. E. Barsach, Zh. Eksp. Teor. Fis. \textbf{69}, 1872 (1975).

\bibitem{Vahtras1992} O. Vahtras \textit{et al}. J. Chem. Phys. \textbf{96}, 6120 (1992).

\bibitem{Raffelt2008} G.G. Raffelt: \textit{Astrophysical Axion Bounds}, Lect. Notes Phys. \textbf{741}, 51–71 (Springer-Verlag Berlin 2008).

\bibitem{Adler2002} S. Adler \textit{et al.}, Phys. Lett. B  \textbf{537}, 211 (2002).

\bibitem{Georgi1986} H. Georgi, D. B. Kaplan, and L. Randall, Physics Letters, \textbf{169B} 73 (1986).

\bibitem{Vaara2002} J. Vaara, J. Jokisaari, E. Wasylishen, and D. L. Bryce, Prog. Nucl. Magn. Res. Spec. \textbf{41}, 233 (2002).

\bibitem{Ledbetter2009} M. P. Ledbetter \textit{et al}. J. Magn. Res. \textbf{199}, 25 (2009).

\bibitem{Bagno2003} A. Bagno \& G. Saielli \textit{Chem. Eur. J.} \textbf{9}, 1486-1495 (2003).

\bibitem{Ledbetter2012} M. P. Ledbetter, A. Bagno, G. Saielli, N. Tran, and M. V. Romalis, arXiv:1112.5644 (2012).

\bibitem{Pecul2008} T. Helgaker, M. Jaszu\'nski, M. Pecul, Prog. Nucl. Magn. Res. Spec. \textbf{53}, 249 (2008).

\bibitem{Hiza1981} M. J. Hiza, Fluid Phase Equilibria \textbf{6}, 203 (1981).

\bibitem{Heckman2003} J. J. Heckman, M. P. Ledbetter, \& M. V. Romalis,  Phys. Rev. Lett. \textbf{91}, 067601 (2003).



\end{thebibliography}
\end{document}